\documentclass[conference]{IEEEtran}
\usepackage{cite}
\usepackage{fancyhdr}
\usepackage{comment}
\usepackage{amsmath,amssymb,amsfonts}
\usepackage{algorithmic}
\usepackage{graphicx}
\usepackage{textcomp}
\usepackage{xcolor}
\usepackage{pifont}
\usepackage{ulem}
\def\BibTeX{{\rm B\kern-.05em{\sc i\kern-.025em b}\kern-.08em
 T\kern-.1667em\lower.7ex\hbox{E}\kern-.125emX}}
\usepackage{balance}
\usepackage{enumitem}
\setlist[enumerate]{leftmargin=*}

\begin{document}

\title{\LARGE 
AutoLock: Automatic Design of Logic Locking with \\Evolutionary Computation\vspace{-0.05in}\\{\large DSN-2023: Doctoral Forum Submission}\vspace{-0.18in}} 

 \author{\IEEEauthorblockN{Zeng Wang\IEEEauthorrefmark{1},
 Lilas Alrahis\IEEEauthorrefmark{2},
 Dominik Sisejkovic\IEEEauthorrefmark{3},
 and Ozgur Sinanoglu\IEEEauthorrefmark{2}
}

\IEEEauthorblockA{
\IEEEauthorrefmark{1}Tandon School of Engineering, New York University, Brooklyn \\
\IEEEauthorrefmark{2}Center for Cyber Security, New York University Abu Dhabi, UAE \\ \IEEEauthorrefmark{3} Security, Privacy, and Safety Research Group, Corporate Research, Robert Bosch GmbH, Germany}
  \{zw3464, lma387, os22\}@nyu.edu, dominik.sisejkovic@de.bosch.com \vspace{-0.18in}
 }

\maketitle

% \begin{comment}
\renewcommand{\headrulewidth}{0.0pt}
\thispagestyle{fancy}
\lhead{}
\rhead{}
\chead{\copyright~2023 IEEE/IFIP.
This is the author's version of the work.
The definitive Version of Record is published in the IEEE/IFIP International Conference on Dependable Systems and Networks (DSN 2023).}
\cfoot{}

% \end{comment}

\pagestyle{plain}

\begin{abstract}
Logic locking protects the integrity of hardware designs throughout the integrated circuit supply chain. However, recent machine learning (ML)-based attacks have challenged its fundamental security, initiating the requirement for the design of learning-resilient locking policies. A promising ML-resilient locking mechanism hides within multiplexer-based locking. Nevertheless, recent attacks have successfully breached these state-of-the-art locking schemes, making it ever more complex to manually design policies that are resilient to all existing attacks. In this project, for the first time, we propose the automatic design exploration of logic locking with evolutionary computation (EC)---a set of versatile black-box optimization heuristics inspired by evolutionary mechanisms. The project will evaluate the performance of EC-designed logic locking against various types of attacks, starting with the latest ML-based link prediction. Additionally, the project will provide guidelines and best practices for using EC-based logic locking in practical applications.

\color{black}
\end{abstract}

\IEEEoverridecommandlockouts

\IEEEpeerreviewmaketitle

% ===================
% # I. Introduction #
% ===================
  \vspace{-2pt}
\section{Introduction}
The contemporary integrated circuit (IC) supply chain has become increasingly complex with the involvement of multiple parties, exposing the genuine hardware design to a range of attacks, including intellectual property (IP) piracy, reverse engineering, counterfeiting, and hardware Trojans~\cite{rostami2014primer, bhunia2018hardware}. These hardware-based attacks can result in financial losses for design companies or IP owners and pose potential risks to end users. For instance, in 2017, infringement of design IP was estimated to cause financial losses of \$225-600 billion for US companies, according to a report by Blair \textit{et al.}~\cite{blair2017update}.

To address these challenges, various countermeasures have been proposed, among which logic locking (LL) has emerged as a promising method for IP protection~\cite{yasin2015improving}. LL is implemented by inserting key-controlled gates, known as \textit{key-gates}, into the \textit{netlist}.\footnote{%A netlist is a description of a digital circuit at the gate level, where each gate is represented as a logical element with a specific type. It specifies how the gates are interconnected to form the complete circuit.
A netlist is a description of a digital circuit at the gate level, specifying gate types and interconnections.} A correct key preserves the original circuit behavior, while incorrect keys lead to erroneous outputs.

Over the past decade, researchers have evaluated the security of LL by developing attacks and defenses. Initially, attacks were primarily based on SAT solvers, but as machine learning (ML) has emerged, the focus has gradually shifted towards incorporating ML in LL attacks and developing corresponding countermeasures~\cite{9606979}. For instance, the \textit{SnapShot} attack~\cite{sisejkovic2021challenging} automatically designs neural network architectures to effectively classify the key-gate sub-circuits to deduce the correct key. Recent ML-based attacks such as \textit{GNNUnlock}~\cite{alrahis2021gnnunlock} and \textit{OMLA}~\cite{alrahis2021omla} demonstrate the robust capability of graph neural networks (GNNs) in distinguishing locked key-gate nodes and deciphering the keys. These efficient ML-based attacks have prompted the design of ML-resilient locking schemes, as showcased in \textit{UNSAIL}~\cite{alrahis2021unsail} and \textit{D-MUX}~\cite{sisejkovic2021deceptive}, which aim to mitigate the vulnerabilities of traditional LL schemes.
However, the state-of-the-art ML-based attack \textit{MuxLink}~\cite{alrahis2022muxlink} extended the attack surface to link prediction, which successfully compromised D-MUX and subsequent symmetric MUX-based locking schemes. \textit{This highlights the need for further research in developing more robust and resilient LL techniques that can defend against these evolving ML-based attacks.}

Despite the continuous development of the LL landscape~\cite{yasin2020trustworthy, sisejkovic2022Book}, it has become increasingly complex to \textit{manually} adapt LL schemes with every new attack. A manual LL design has time and again led to overlooking critical vulnerabilities. Therefore, \textbf{in this project}, we propose \textit{AutoLock}; an automatic approach to designing LL schemes that utilizes the heuristic optimization capabilities of genetic algorithms (GA).\footnote{GAs are a type of heuristic algorithm in evolutionary computation~\cite{eiben2015introduction}.} As GAs are based on a black-box optimization, we aim to design a framework that can automatically explore the design of resilient LL based on user-defined security objectives.

\noindent\textbf{Overcoming Challenges and Realizing Potential:} \textit{Despite the potential benefits of using GAs to automate LL design, such as increased efficiency and security, this approach has not been explored.} Challenges arise due to the complexity of the design space, the need to optimize multiple criteria, and the evolving nature of attacks. Moreover, incomplete understanding of the underlying mechanisms and best practices for LL makes it difficult to formulate a suitable fitness function for GAs. \color{black}However, our research project aims to address these challenges and lay the foundation for the future use of GAs in automating LL design. 
The initial goal is to integrate the MuxLink attack with a GA to guide the search process toward resilient D-MUX-based locking. The overall objective of the study is to explore the capabilities, requirements, and limitations of a GA-based design of resilient LL.

% =======================================================
% # II. MuxLink prediction stragety #
% =======================================================

\begin{figure*}[ht]
 \centering
 \includegraphics[width=0.9\textwidth]{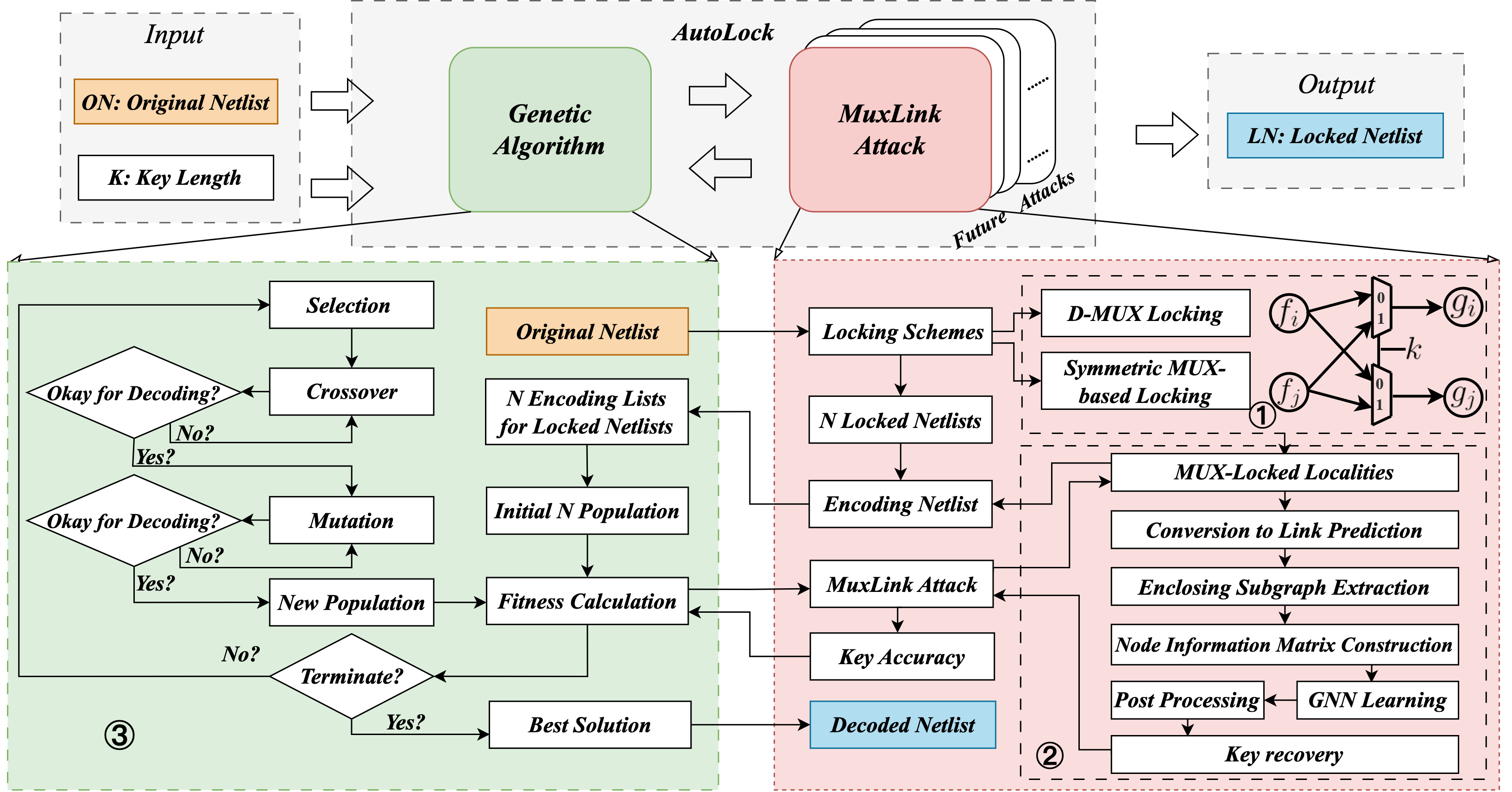}
   \vspace{-5pt}
 \caption{\small Workflow of the automatic design for MUX-based LL.}
  \vspace{-15pt}
 \label{fig:MUX-GA}
\end{figure*}
  \vspace{-5pt}
\section{Proposed AutoLock Obfuscation Approach}

\textcolor{black}{D-MUX inserts key-controlled pairs of multiplexers to create additional paths between selected netlist nodes. For example, as shown in Fig.~\ref{fig:MUX-GA}\ding{172}, D-MUX inserts the multiplexers between the initially connected nodes $\{f_i, g_i\}$ and $\{f_j, g_j\}$ thereby creating multiple valid paths. Only a correct key ensures that the correct paths are enabled. The nodes $\{f_i, f_j, g_i, g_j\}$ define a \textit{locality}.\footnote{A locality is a sub-circuit surrounding a key-gate.} Even though D-MUX is resilient against locality-based learning attacks, it ignores the fan-in and fan-out of these localities. MuxLink exploits this vulnerability to determine a correct connection based on the surrounding gates, as shown in Fig.~\ref{fig:MUX-GA}\ding{173}. }

\textbf{GA-MuxLink Integration:}  
AutoLock deploys the GA to automatically refine a locked netlist until it reaches a desired fitness. The input to AutoLock is the original netlist (ON) and desired key length (K). The output is the locked netlist (LN) that fulfills the selected security objectives. The key component of GAs is the \textit{genotype}. The genotype is a representation (encoding) of the optimization problem in a form that can be processed by the GA. The genotype is almost always problem-specific. We defined the genotype in form of a list of pairs $\{f_i, f_j, g_i, g_j, k\}$, where $k$ is the correct key bit. Hence, each element of the genotype uniquely identifies a location to insert a MUX-locked key-gate. As shown in Fig.~\ref{fig:MUX-GA}, we first lock the provided ON with a key of size K. This is repeated N times with random keys. The N resulting encodings are passed to the GA (Fig.~\ref{fig:MUX-GA}\ding{174}) to initialize the population. 
The fitness of each genotype %(solution)
is measured by MuxLink accuracy, where lower accuracy indicates higher fitness. The 
process stops when a set number of iterations or desired fitness is achieved.

The GA utilizes (mainly) three evolutionary operators---selection, crossover, and mutation---to generate better offspring throughout multiple generations. In each generation, the selection determines which parents (encodings) are selected for a crossover. The crossover generates new children by combining the parental genetic material (list of $\{f_i, f_j, g_i, g_j, k\}$). Finally, the new children are mutated by introducing small changes in their encodings. The mutation is important to ensure the exploration of new genetic material. 

\textbf{First Insights:} Currently, the AutoLock tool enables a successful integration between MuxLink and the GA by encoding the original locked netlist into the developed genotype representation. First experimental results (without parameter tuning) indicate the capability of AutoLock to generate locked netlists that \textit{successfully decrease the attack accuracy} by 25 percentage points. This is a strong indication that utilizing GAs for LL design is a promising research direction.

% =============================================
% # IV. Research Plan#
% =============================================
 \vspace{-3pt}
\section{Research Plan}
In our pursuit of automatically designing LL schemes, several research directions require attention in the future:
\begin{enumerate}

 \item[$\circ$] The success of evolutionary search relies on selecting the correct genotype representation. We will evaluate the limitations and opportunities of our selected encoding.

 \item[$\circ$] As the optimization success of the GA depends on the design of the evolutionary operators, we need to take a look at the design of problem-specific operators.

 \item[$\circ$] The GA-based MUX locking scheme proposed in our work uses the MuxLink attack to determine the fitness. While this approach can effectively protect against MuxLink attacks, it does not guarantee resilience against other attack vectors. Thus, there is still a need to evaluate a multi-objective optimization that includes a set of distinct attacks
 \item[$\circ$] The overall objective of this project is to provide the community with AutoLock; a platform for the automatic design of resilient LL. The evolutionary process could provide further insights into mitigating unexplored vulnerabilities.
  \item[$\circ$] Finally, depending on the security objectives, we will explore other techniques out of the evolutionary computation field to better understand what heuristics are more suitable for this form of automation.
 
\end{enumerate}

% ==============
% # REFERENCES #
% ==============

\balance
\bibliographystyle{IEEEtran}
\bibliography{bibliography.bib}

% Generated by IEEEtran.bst, version: 1.14 (2015/08/26)
\begin{thebibliography}{10}
\providecommand{\url}[1]{#1}
\csname url@samestyle\endcsname
\providecommand{\newblock}{\relax}
\providecommand{\bibinfo}[2]{#2}
\providecommand{\BIBentrySTDinterwordspacing}{\spaceskip=0pt\relax}
\providecommand{\BIBentryALTinterwordstretchfactor}{4}
\providecommand{\BIBentryALTinterwordspacing}{\spaceskip=\fontdimen2\font plus
\BIBentryALTinterwordstretchfactor\fontdimen3\font minus
  \fontdimen4\font\relax}
\providecommand{\BIBforeignlanguage}[2]{{%
\expandafter\ifx\csname l@#1\endcsname\relax
\typeout{** WARNING: IEEEtran.bst: No hyphenation pattern has been}%
\typeout{** loaded for the language `#1'. Using the pattern for}%
\typeout{** the default language instead.}%
\else
\language=\csname l@#1\endcsname
\fi
#2}}
\providecommand{\BIBdecl}{\relax}
\BIBdecl

\bibitem{rostami2014primer}
M.~Rostami, F.~Koushanfar, and R.~Karri, ``A primer on hardware security:
  Models, methods, and metrics,'' \emph{Proceedings of the IEEE}, vol. 102,
  no.~8, pp. 1283--1295, 2014.

\bibitem{bhunia2018hardware}
S.~Bhunia and M.~Tehranipoor, ``The hardware trojan war,'' \emph{Cham,
  Switzerland: Springer}, 2018.

\bibitem{blair2017update}
D.~Blair, J.~Huntsman~Jr, C.~Barrett, S.~Gordon, W.~Lynn~III, D.~WinceSmith,
  and M.~Young, ``Update to the {IP} commission report: The report of the
  commission on the theft of american intellectual property,'' \emph{The
  National Bureau of Asian Research}, 2017.

\bibitem{yasin2015improving}
M.~Yasin, J.~J. Rajendran, O.~Sinanoglu, and R.~Karri, ``On improving the
  security of logic locking,'' \emph{IEEE Transactions on Computer-Aided Design
  of Integrated Circuits and Systems}, vol.~35, no.~9, pp. 1411--1424, 2015.

\bibitem{9606979}
D.~Sisejkovic, L.~M. Reimann, E.~Moussavi, F.~Merchant, and R.~Leupers, ``Logic
  locking at the frontiers of machine learning: A survey on developments and
  opportunities,'' in \emph{2021 IFIP/IEEE 29th International Conference on
  Very Large Scale Integration (VLSI-SoC)}, 2021, pp. 1--6.

\bibitem{sisejkovic2021challenging}
D.~Sisejkovic, F.~Merchant, L.~M. Reimann, H.~Srivastava, A.~Hallawa, and
  R.~Leupers, ``Challenging the security of logic locking schemes in the era of
  deep learning: A neuroevolutionary approach,'' \emph{ACM Journal on Emerging
  Technologies in Computing Systems (JETC)}, vol.~17, no.~3, pp. 1--26, 2021.

\bibitem{alrahis2021gnnunlock}
L.~Alrahis, S.~Patnaik, F.~Khalid, M.~A. Hanif, H.~Saleh, M.~Shafique, and
  O.~Sinanoglu, ``Gnnunlock: Graph neural networks-based oracle-less unlocking
  scheme for provably secure logic locking,'' in \emph{2021 Design, Automation
  \& Test in Europe Conference \& Exhibition (DATE)}.\hskip 1em plus 0.5em
  minus 0.4em\relax IEEE, 2021, pp. 780--785.

\bibitem{alrahis2021omla}
L.~Alrahis, S.~Patnaik, M.~Shafique, and O.~Sinanoglu, ``Omla: An oracle-less
  machine learning-based attack on logic locking,'' \emph{IEEE Transactions on
  Circuits and Systems II: Express Briefs}, vol.~69, no.~3, pp. 1602--1606,
  2021.

\bibitem{alrahis2021unsail}
L.~Alrahis, S.~Patnaik, J.~Knechtel, H.~Saleh, B.~Mohammad, M.~Al-Qutayri, and
  O.~Sinanoglu, ``Unsail: Thwarting oracle-less machine learning attacks on
  logic locking,'' \emph{IEEE Transactions on Information Forensics and
  Security}, vol.~16, pp. 2508--2523, 2021.

\bibitem{sisejkovic2021deceptive}
D.~Sisejkovic, F.~Merchant, L.~M. Reimann, and R.~Leupers, ``Deceptive logic
  locking for hardware integrity protection against machine learning attacks,''
  \emph{IEEE Transactions on Computer-Aided Design of Integrated Circuits and
  Systems}, vol.~41, no.~6, pp. 1716--1729, 2021.

\bibitem{alrahis2022muxlink}
L.~Alrahis, S.~Patnaik, M.~Shafique, and O.~Sinanoglu, ``Muxlink: Circumventing
  learning-resilient mux-locking using graph neural network-based link
  prediction,'' in \emph{2022 Design, Automation \& Test in Europe Conference
  \& Exhibition (DATE)}.\hskip 1em plus 0.5em minus 0.4em\relax IEEE, 2022, pp.
  694--699.

\bibitem{yasin2020trustworthy}
M.~Yasin, J.~J. Rajendran, and O.~Sinanoglu, \emph{Trustworthy Hardware Design:
  Combinational Logic Locking Techniques}.\hskip 1em plus 0.5em minus
  0.4em\relax Springer International Publishing, 2020.

\bibitem{sisejkovic2022Book}
D.~Sisejkovic and R.~Leupers, \emph{Logic Locking: A Practical Approach to
  Secure Hardware}.\hskip 1em plus 0.5em minus 0.4em\relax Springer, 2023.

\bibitem{eiben2015introduction}
A.~E. Eiben and J.~E. Smith, \emph{Introduction to evolutionary
  computing}.\hskip 1em plus 0.5em minus 0.4em\relax Springer, 2015.

\end{thebibliography}

\end{document}